# Autonomous materials search using machine learning and ab initio calculations for L1$_0$-FePt-based quaternary alloys

Yuma Iwasaki[a*], Daisuke Ogawa[b], Masato Kotsugi[c], Yukiko K. Takahashi[b]

[a] *Center for Basic Research on Materials (CBRM), National Institute for Materials Science (NIMS), Tsukuba, Japan*

[b] *Research Center for Magnetic and Spintronic Materials (CMSM), National Institute for Materials Science (NIMS), Tsukuba, Japan*

[c] *Department of Materials Science and Technology, Tokyo University of Science, Tokyo, Japan*

ORCID

Yuma Iwasaki: 0000-0002-7117-277X

Daisuke Ogawa: 0000-0002-4373-6435

Masato Kotsugi: 0000-0002-4841-1808

Yukiko Takahashi: 0000-0001-9197-7236

*Corresponding author email: IWASAKI.Yuma@nims.go.jp

# Autonomous materials search using machine learning and ab initio calculations for $L1_0$-FePt-based quaternary alloys

The efficient exploration of expansive material spaces remains a significant challenge in materials science. To address this issue, autonomous material search methods that combine machine learning with ab initio calculations have emerged as a promising solution. These approaches offer a systematic and rapid means of discovering new materials, particularly when the material space is too large. This requirement is particularly important in the development of $L1_0$-structured alloys as magnetic recording media. These materials require a high magnetic moment ($M$) and magnetocrystalline anisotropy energy ($E_{MCA}$) to satisfy the demands of next-generation data storage technologies. Although autonomous search methods have been successfully applied to various material systems, quaternary $L1_0$ alloys with optimized magnetic properties remain an open and underexplored frontier. In this study, we present a simulation-based autonomous search method aimed at identifying quaternary $L1_0$ alloys with enhanced $M$ and $E_{MCA}$ values. Over a continuous 100-day search, our system suggested the FeMnPtEr alloy system as a promising candidate, exhibiting superior values for both $M$ and $E_{MCA}$. Although further experimental validation is required, this study underscores the potential of autonomous search methods to accelerate the discovery of advanced materials.

## 1. Introduction

The exploration of material spaces has significantly expanded in recent years. Traditionally, researchers navigate these vast material spaces through iterative cycles of material synthesis, property measurements, and subsequent analyses. However, this conventional approach is time-consuming and inadequate for comprehensively exploring expanded material spaces. To overcome these limitations, machine learning has been incorporated into automated material search methodologies. These methodologies can be broadly classified into two main categories: autonomous searches using robotics and material simulations. In the first approach, robots are employed to automate tasks, such as material synthesis and property measurement. Machine learning is then used to analyze the collected data to suggest materials and optimal process conditions for subsequent synthesis cycles. This closed-loop system, driven by robotic synthesis and measurement, enables autonomous material searches. The development of various autonomous material search robots has been established as an effective strategy in material development [1–12]. In the second approach, simulations such as ab initio calculations are employed to obtain material property data, which are then analyzed using machine learning to guide future material designs in terms of structure and composition. This iterative process forms a closed loop, allowing the autonomous exploration of material spaces in a computational environment, offering flexibility, versatility, and broad applicability [13–22].

In this study, we explored a quaternary alloy space based on $L1_0$-FePt using an autonomous material search method combining ab initio calculations and machine learning. $L1_0$-FePt is considered to be a strong candidate for next-generation magnetic recording materials because of its high magnetocrystalline anisotropy, magnetic moment, corrosion resistance, and oxidation resistance [23-26]. However, recent trends toward miniaturization require new materials with even larger magnetic moments and magnetocrystalline anisotropy. By adding third and fourth elements to $L1_0$-FePt, it may be possible to enhance both the magnetic moment and magnetocrystalline anisotropy, making it a promising candidate for magnetic recording media. However, given the wide variety of possible additive elements and their concentrations, the number of potential combinations is large. It is practically infeasible to synthesize and evaluate all the potential candidate materials within this extensive material space. Moreover, evaluating magnetocrystalline anisotropy through ab initio calculations is computationally expensive, making exhaustive searches via this method impractical.

Therefore, proposing material compositions that exhibit both high magnetic moments and magnetocrystalline anisotropy energy through autonomous material search methods is a promising approach for successful synthesis and practical applications.

## 2. Methods

The material space was defined as quaternary materials based on L1$_0$-FePt.

$$Fe_{1-x}X_xPt_{1-y}Y_y \tag{1}$$

where *X* and *Y* are defined as

$$X = \{Al, Si, Sc, Ti, V, Cr, Mn, Fe, Co, Ni, Cu, Zn, Ag\} \tag{2}$$

$$Y = \begin{Bmatrix} Ga, Ge, Y, Zr, Nb, Mo, Ru, Rh, Pd, Cd, In, Sn, Sb, \\ La, Ce, Pr, Nd, Sm, Eu, Gd, Tb, Dy, Ho, Er, Tm, Yb, Lu, \\ Hf, Ta, W, Re, Os, Ir, Pt, Au, Tl, Pb, Bi \end{Bmatrix} \tag{3}$$

The assignment of each atom to either X or Y was determined based on the total energy obtained from DFT calculations. The composition *x* and *y* at each site was incremented by 0.01 between 0 to 0.2.

$$x, y = \{0, 0.01, 0.02, 0.03 \ldots, 0.2\} \tag{4}$$

Approximately 200,000 potential material configurations exist in a specified material space. The challenge of performing ab initio calculations for all candidates was alleviated by using sequential calculations guided by machine learning to efficiently navigate the material space.

An autonomous search simultaneously maximizes two key indicators. The first indicator is the total magnetic moment (*M*), which is proportional to the saturation magnetization. In recent years, as magnetic recording media have become more densely packed, the reduction in magnetic signals per domain has emerged as a significant issue. Therefore, a higher magnetic moment is preferable.

The second indicator is the magnetocrystalline anisotropy energy ($E_{MCA}$), which is calculated as the energy difference between the easy [001] and hard [100] magnetization axes. A higher magnetocrystalline anisotropy energy is also desirable to ensure the stable retention of the recorded magnetic information.

The methodology used for the autonomous material search is illustrated in Figure 1. It includes sequential ab initio calculations and machine learning. The ab initio phase calculates $M$ and $E_{MCA}$ based on the compositions recommended by the machine learning phase. The data accumulated from these calculations guided the composition choices for subsequent ab initio calculations.

The ab initio calculations used Green's function-based density functional theory using the Korringa–Kohn–Rostoker coherent potential approximation (KKR–CPA) method, implemented in AkaiKKR software [27]. CPA allows the accurate simulation of alloy systems, particularly for multi-element disordered phases [28–31]. Further details of the ab initio calculations are provided in the Supplementary Materials (S1).

The machine-learning phase used Bayesian optimization [32] and an autoencoder [33] to identify the material compositions for subsequent KKR–CPA calculations based on the accumulated $M$ and $E_{MCA}$ data. The material space (descriptors) was designed based on previous reports [19,21,22]. The composition and Magpie descriptor vectors [34] were compressed into a 30-dimensional latent vector representing the unexplored material space. Additional details are provided in Supplementary Materials (S2). This phase integrated KKR-CPA with multi-objective Bayesian optimization using $M$ and $E_{MCA}$ as objective variables, and the latent vectors generated by the autoencoder as explanatory variables in the Gaussian process regression model. The upper confidence bound (UCB) was calculated as the acquisition function for each material [35]. The candidate material with the highest Pareto hypervolume based on the UCB value and training data (observed $M$ and $E_{MCA}$) was selected for subsequent KKR–CPA calculations. This iterative approach facilitated the autonomous exploration of materials exhibiting high $M$ and $E_{MCA}$. Further details are provided in the Supplementary Materials (S3).

## 3. Results and Discussion

The developed autonomous search system operated continuously for approximately 100 days. Figure 2 shows a two-dimensional plot of the explored materials illustrating their $M$ and $E_{MCA}$ characteristics. The white circles represent the initial data precalculated for $Fe_{0.99}X_{0.01}Pt_{1.00}$ and $Fe_{1.00}Pt_{0.99}Y_{0.01}$, indicating their respective $M$ and $E_{MCA}$ values. Elements X and Y are given by Equations (1) and (2), respectively. The black circles indicate the $M$ and $E_{MCA}$ values of the materials explored using the autonomous search system. Several materials with higher $M$ and $E_{MCA}$ than the initial values were proposed. The system also sometimes explored materials with lower $M$ and $E_{MCA}$ compared than the initial data. It is important to learn from both good (high $M$ and $E_{MCA}$) and poor data (low $M$ and $E_{MCA}$) to improve the accuracy of machine learning models. Using the UCB strategy, this autonomous search system also explored materials with low $M$ and $E_{MCA}$ at an appropriate rate, thereby improving its machine learning model.

The autonomous search system frequently had been exploring $Fe_{1-x}Mn_xPt_{1-y}Er_y$ alloys as promising candidates with high values for both $M$ and $E_{MCA}$. Therefore, a comprehensive calculation of the $Fe_{1-x}Mn_xPt_{1-y}Er_y$ system was performed using the KKR-CPA method. Figure 3(a) shows the composition dependence of $M$ in the $Fe_{1-x}Mn_xPt_{1-y}Er_y$ system. The addition of Mn and Er to FePt increases the magnetic moment. Figure 3(b) shows the composition dependence of $E_{MCA}$ in the same system. Similar to the behavior of $M$, the addition of Mn and Er to FePt increased $E_{MCA}$. Because the addition of Mn and Er improved both $M$ and $E_{MCA}$, the autonomous search system considered the $Fe_{1-x}Mn_xPt_{1-y}Er_y$ system to be a promising material and focused its exploration on this system.

Figure 4(a) shows the compositional dependence of the lattice constant of the $Fe_{1-x}Mn_xPt_{1-y}Er_y$ system. The addition of Mn slightly increases the lattice constant, whereas the addition of Er significantly increases it. Figure 4(b) shows the compositional dependence of the c/a ratio of the $Fe_{1-x}Mn_xPt_{1-y}Er_y$ system. The addition of Mn slightly increases the c/a ratio, while the addition of Er significantly increases it, leading to a near-cubic structure (c/a ≈ 1.0) when the Er concentration is y = 0.2.

To investigate the origin of the increase in $M$ caused by the addition of Mn and Er, as shown in Figure 3(a), the composition dependence of the local magnetic moments for each element in the $Fe_{1-x}Mn_xPt_{1-y}Er_y$ system was calculated. Figure 5(a) shows the composition dependence of the local magnetic moment of Fe ($M_{Fe}^T$). The addition of Mn enhanced the local magnetic moment of Fe, whereas the addition of Er reduced it. Figure 5(b) shows the compositional dependence of the local magnetic moment of Mn ($M_{Mn}^T$). As the Mn concentration increased, the local magnetic moment of Mn also increased; however, the addition of Er tended to reduce it. The increase in the local magnetic moment of Fe, which is the primary contributor to the total magnetic moment, combined with the fact that Mn possesses a larger local magnetic moment than Fe, explains the slight increase in the overall magnetic moment along the x-axis (Mn concentration; Fig. 3(a)).

Figure 5(c) shows the compositional dependence of the local magnetic moment of Pt ($M_{Pt}^T$). The addition of Mn and Er reduces the local magnetic moment of Pt. Figure 5(d) shows the composition dependence of the local magnetic moment of Er ($M_{Er}^T$). When the Er concentration was low (< 10 at%), the addition of Mn reduced the local magnetic moment of Er; however, when the Er concentration was high (> 10 at%), the addition of Mn increased the local magnetic moment of Er. Meanwhile, increasing the Er concentration reduce the Er local magnetic moment. Because the Er magnetic moment is larger than those of Fe and Mn, the overall increase in the magnetic moment along the y-axis (Er amount) in Figure 3(a) is primarily due to the replacement of Pt, which has a smaller local magnetic moment, with Er, which has a larger local magnetic moment. However, this result was based on ab initio calculations at absolute zero. The magnetic moment of lanthanides, such as Er, is generally large at absolute zero but decreases significantly at room temperature. Therefore, even if a material with Er added to FePt is synthesized and its magnetization is measured at room temperature, a significant increase in magnetization along the y-axis (Er amount) of Figure 3(a) may not be observed.

To investigate the enhancement of $E_{MCA}$ caused by the addition of Er, as shown in Figure 3(b), the composition dependence of the spin and orbital magnetic moments for each element in the $Fe_{1-x}Mn_xPt_{1-y}Er_y$ system was calculated. Magnetocrystalline anisotropy is believed to originate from spin-orbit interactions and orbital magnetic moments [36]. Figures 6(a-d) show the spin magnetic moments of Fe, Mn, Pt, and Er, respectively, in the $Fe_{1-x}Mn_xPt_{1-y}Er_y$ system. These trends approximately correspond to those of the total magnetic moments shown in Figures 5(a-d).

Figures 7(a)–(d) show the orbital magnetic moments of Fe, Mn, Pt, and Er in the $Fe_{1-x}Mn_xPt_{1-y}Er_y$ system. The addition of Mn slightly reduces the orbital magnetic

moments of Fe, Mn, Pt, and Er. In contrast, the addition of Er increases the orbital magnetic moments of Fe, Mn, and Pt. Although the orbital magnetic moment of Er decreases slightly with the addition of Er, the orbital magnetic moment of Er was still much larger than those of Fe, Mn, and Pt. Therefore, the significant magnetocrystalline anisotropy observed in the $Fe_{1-x}Mn_xPt_{1-y}Er_y$ system can be attributed to the large orbital magnetic moment of Er and the increase in the orbital magnetic moments of Fe, Mn, and Pt due to the addition of Er.

More specifically, magnetocrystalline anisotropy is explained by the anisotropy of the orbital magnetic moment, $\Delta M_o^T$ [18,36]. Figure 8 illustrates $\Delta M_o^T$ for the $Fe_{1-x}Mn_xPt_{1-y}Er_y$ system. $\Delta M_o^T$ is defined as the difference in total orbital moments between the magnetization directions [001] and [100] [18,36]. The addition of Mn and Er is found to enhance $\Delta M_o^T$, which is considered to be the origin of the large magnetocrystalline anisotropy.

We successfully used autonomous exploration methods to propose new $L1_0$-based quaternary alloys $Fe_{1-x}Mn_xPt_{1-y}Er_y$ with high $M$ and $E_{MCA}$. However, these newly proposed alloy materials are speculative predictions derived from ab initio calculations, and the accuracy of $M$ and $E_{MCA}$ predictions has not been verified. Moreover, the stability of the $L1_0$ structures formed by the elements in these alloys is uncertain. Therefore, further experimental and theoretical investigations on the proposed alloy materials are required.

## 4. Conclusions

Simulation-based autonomous material search methods are highly effective for exploring the extensive alloy space of quaternary $L1_0$ materials, focusing on achieving a high magnetic moment ($M$) and magnetocrystalline anisotropy energy ($E_{MCA}$). By integrating machine learning and ab initio calculations, we demonstrated a system that efficiently navigated the complex material landscape and ultimately proposed a new material, $Fe_{1-x}Mn_xPt_{1-y}Er_y$, with exceptional $M$ and $E_{MCA}$ values. This autonomous search approach demonstrates versatility and adaptability across various material systems and properties and serves as a powerful tool capable of accelerating advancements in material discovery.

Funding details: This study was supported by JST-CREST under Grant JPMJCR21O1 and the Data Creation and Utilization-Type Material Research and Development Project (Digital Transformation Initiative Center for Magnetic Materials) under Grant JPMXP1122715503.

Disclosure statement: The author reports there are no competing interests to declare.

Data availability statement: The data supporting the results of this research are available from the corresponding author upon reasonable request.

Acknowledgements: We thank Y. Miura, K. Masuda, G. Xing, K. Sodeyama, Y. Sakuraba, and Y. Sasaki at the National Institute for Materials Science, and Y. Igarashi at Tsukuba University for their valuable discussions. This study was supported by JST-CREST (Grant No. JPMJCR21O1) and the Data Creation and Utilization-Type Material Research and Development Project (Digital Transformation Initiative Center for Magnetic Materials) under Grant No. JPMXP1122715503.

**Figures and captions**

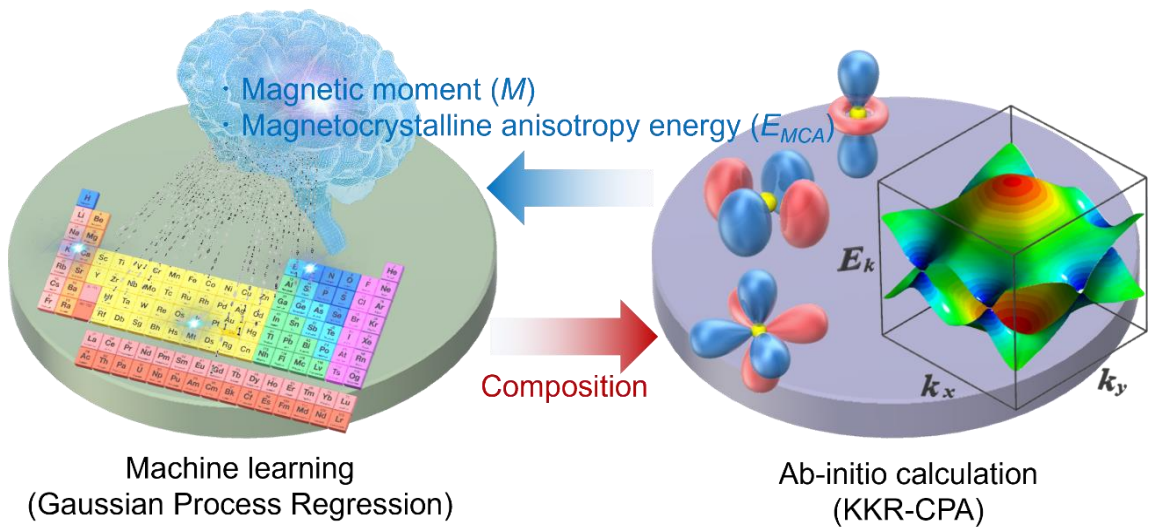

Figure 1. Overview of an autonomous material search system; the *ab initio* calculation phase computes magnetic moment (*M*) and magneto crystalline anisotropy energy ($E_{MCA}$) using crystal structure and composition data determined in the machine learning phase; the machine learning phase generates compositional information for subsequent *ab initio* calculations. This method is based on previous research [19,21,22].

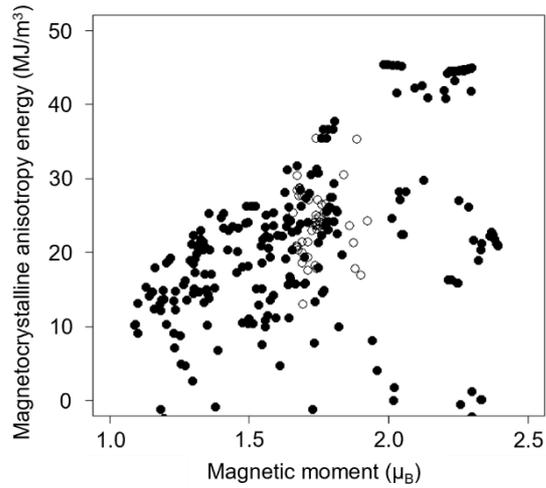

Figure 2. Results of the autonomous search for materials with high magnetic moment ($M$) and magnetocrystalline anisotropy energy ($E_{MCA}$). The white circles represent the initial data, pre-calculated for $Fe_{0.99}X_{0.01}Pt_{1.00}$ and $Fe_{1.00}Pt_{0.99}Y_{0.01}$, showing the respective $M$ and $E_{MCA}$ values. The black circles indicate the $M$ and $E_{MCA}$ values of the materials explored by the autonomous search system.

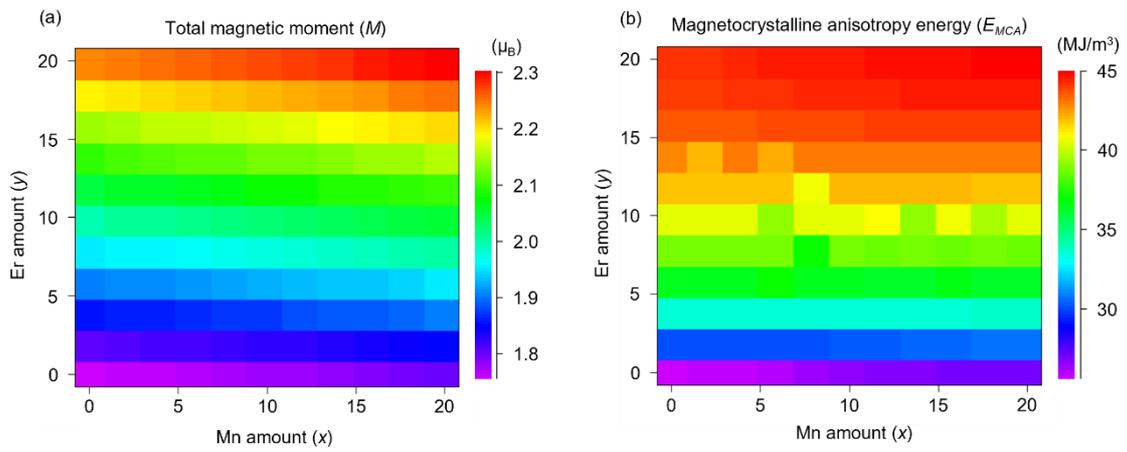

Figure 3. Composition dependence of the **(a)** total magnetic moment (*M*) and **(b)** magnetocrystalline anisotropy energy ($E_{MCA}$) in the $Fe_{1-x}Mn_xPt_{1-y}Er_y$ system.

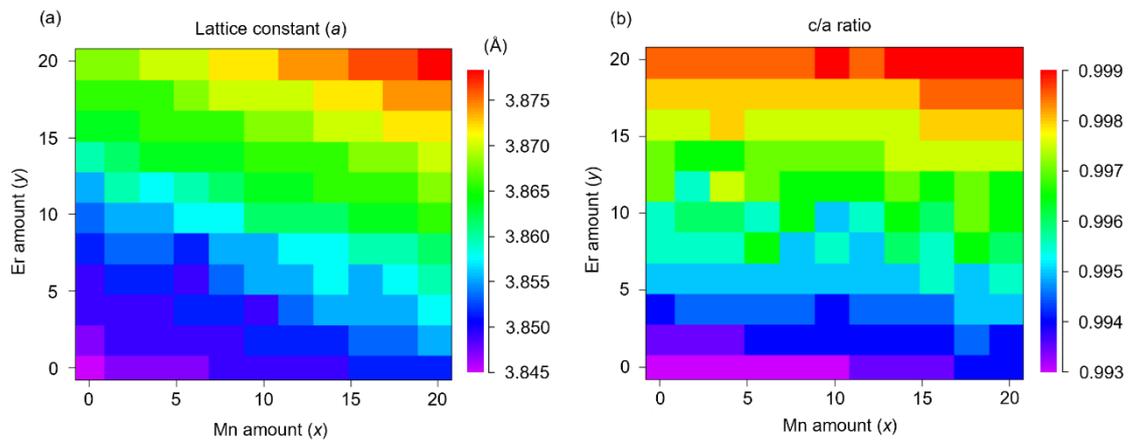

Figure 4. Composition dependence of the (a) lattice constant and (b) c/a ratio in the $Fe_{1-x}Mn_xPt_{1-y}Er_y$ system.

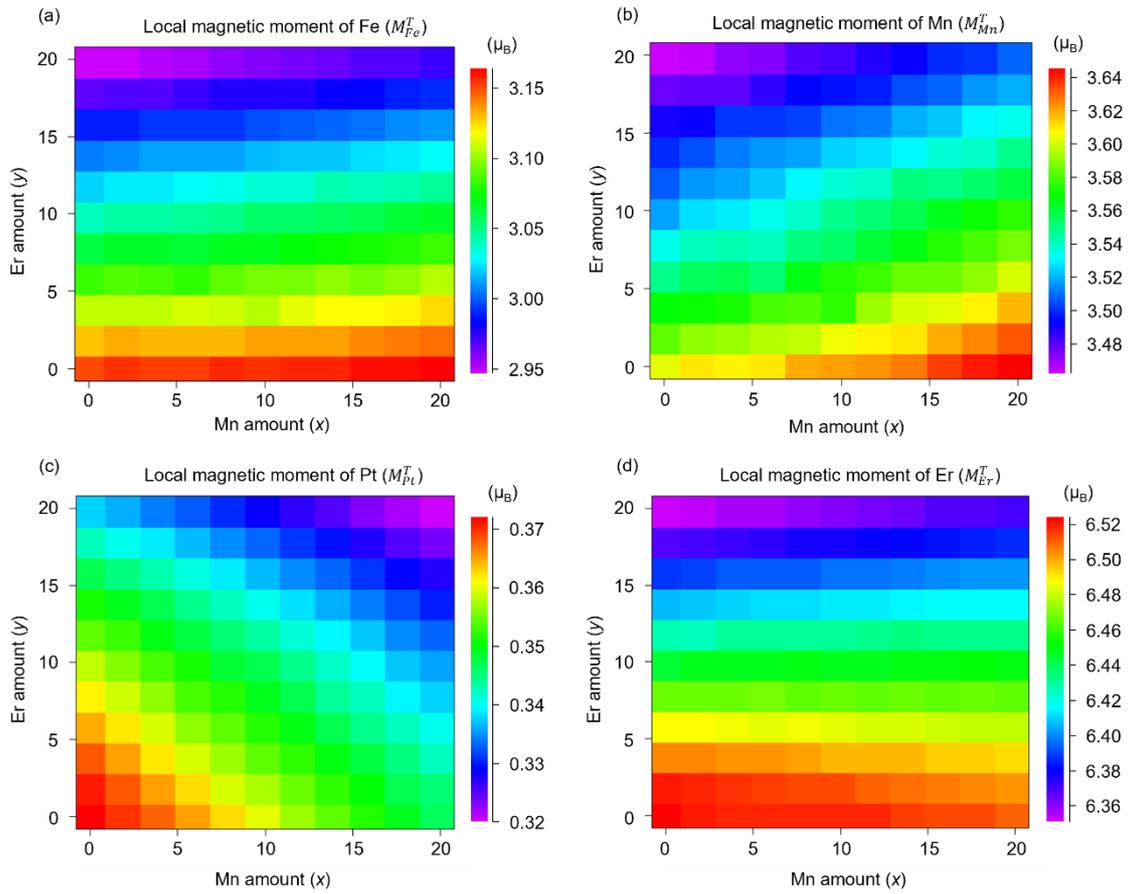

Figure 5. Composition dependence of the **(a)** local magnetic moment of Fe ($M_{Fe}^{T}$), **(b)** local magnetic moment of Mn ($M_{Mn}^{T}$), **(c)** local magnetic moment of Pt ($M_{Pt}^{T}$), and **(d)** local magnetic moment of Er ($M_{Er}^{T}$) in the $Fe_{1-x}Mn_xPt_{1-y}Er_y$ system.

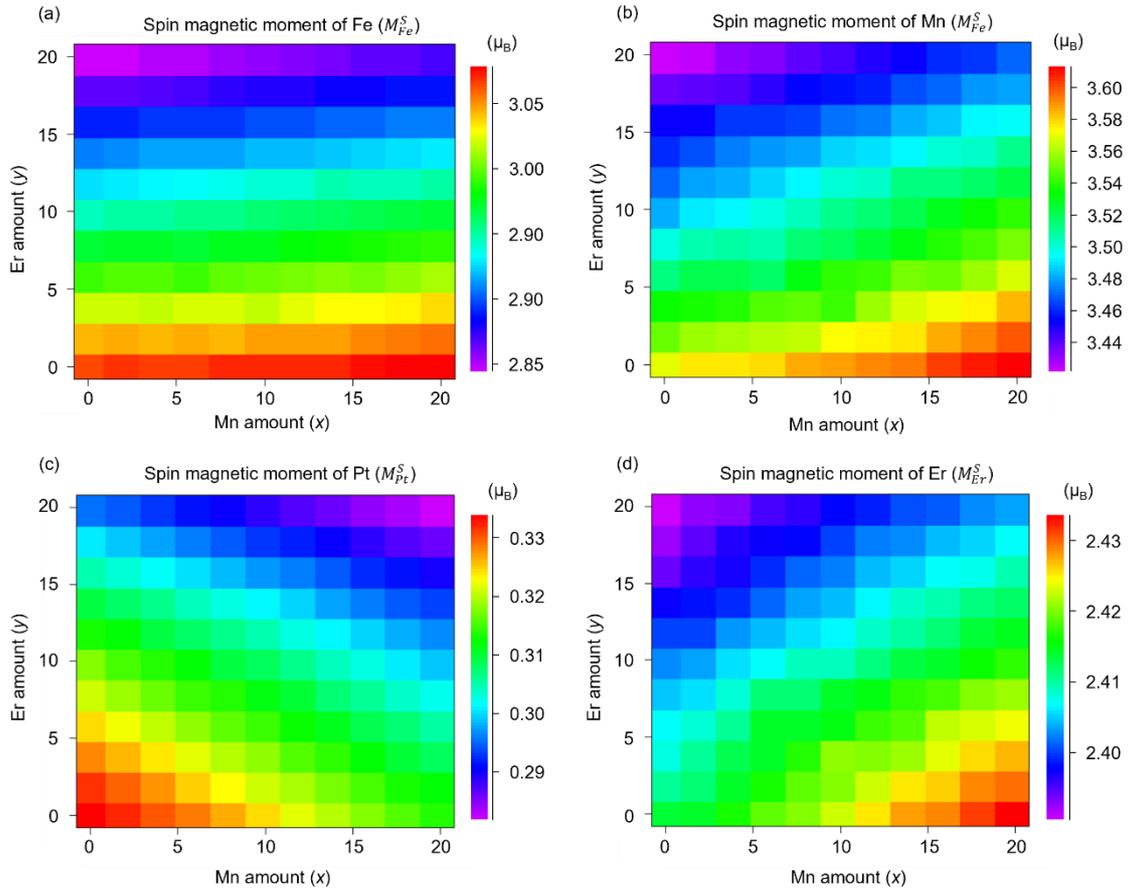

Figure 6. Composition dependence of the **(a)** spin magnetic moment of Fe ($M_{Fe}^S$), **(b)** spin magnetic moment of Mn ($M_{Mn}^S$), **(c)** spin magnetic moment of Pt ($M_{Pt}^S$), and **(d)** spin magnetic moment of Er ($M_{Er}^S$) in the $Fe_{1-x}Mn_xPt_{1-y}Er_y$ system.

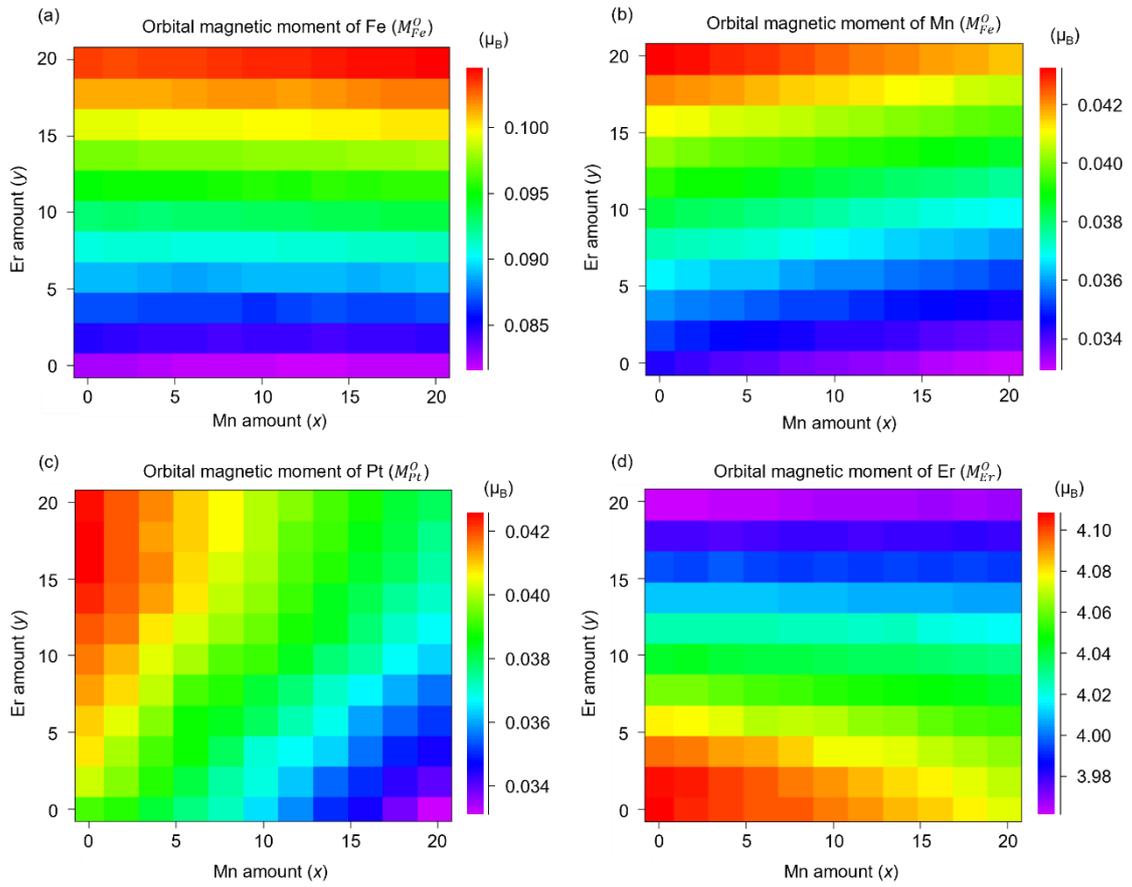

Figure 7. Composition dependence of the **(a)** orbital magnetic moment of Fe ($M_{Fe}^{O}$), **(b)** orbital magnetic moment of Mn ($M_{Mn}^{O}$), **(c)** orbital magnetic moment of Pt ($M_{Pt}^{O}$), and **(d)** orbital magnetic moment of Er ($M_{Er}^{O}$) in the Fe$_{1-x}$Mn$_x$Pt$_{1-y}$Er$_y$ system.

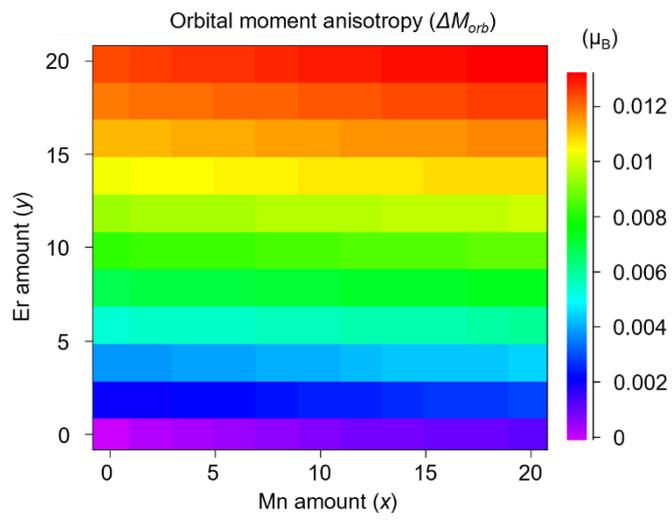

Figure 8. Composition dependence of the orbital moment anisotropy ($\Delta M_o^T$) in the Fe$_{1-x}$Mn$_x$Pt$_{1-y}$Er$_y$ system.